\documentclass[aps,prd, twocolumn]{revtex4}
\usepackage{graphicx}    %  in the preamble
\usepackage{amsmath}

\begin{document}
\title{Integral representation of the cosmic microwave background spectrum}

\author{Moorad Alexanian}
\email[]{alexanian@uncw.edu}
\affiliation{Department of Physics and Physical Oceanography\\
University of North Carolina Wilmington\\ Wilmington, NC
28403-5606\\}

\date{\today}

\begin{abstract}
We use an integral representation for nonthermal radiation, which is bounded from below and above, to describe the spectrum of the cosmic microwave background (CMB). The upper bound is given by the Rayleigh-Jeans law with a temperature $T_{RJ}$ that can be determined by the absorption signal of 21 cm photons, where $T_{RJ}$ represents the equilibrium temperature of photons in the RJ tail. If $T_{RJ}>T_{CMB}$, then the lower bound allows us to conclude that photons, additional to the remnant of the Big Bang, are needed to explain the present CMB. These constraints are additional to other cosmological or astrophysical constraints in the study of the distortions of the CMB brought about by new physics particles or fields.
\end{abstract}

\maketitle {}

\section{Introduction}

The lambda cold dark matter ($\Lambda$CDM) model has been established as the standard cosmological model to describe the expansion history and the growth of the large-scale structure of the Universe \cite{KA18}. Assuming the $\Lambda$CDM model \cite{Hikage}, cosmological parameters have been measured within percent-level uncertainties by a combination of observations such as the cosmic microwave background (CMB) experiments \cite{Hinshaw,Aghanim}. Additional measurements include Type Ia supernovae and baryon acoustic oscillations (See references in \cite{Hikage}). Despite the success of the model, we are challenged by a fundamental lack of physical understanding of the main components of the universe, dark matter, and cosmological constant $\Lambda$ or more generally dark energy. In order to understand these dark components, it is of great importance to test the $\Lambda$CDM model at high precision using a variety of cosmological probes \cite{Hikage}.

Recently, Pospelov, \emph{et al}. \cite{PPRU18} consider modifications of the cosmic microwave background (CMB), within its Rayleigh-Jeans (RJ) end of the spectrum, $\omega\ll T_{CMB}$, owing to dark matter (DM) and the interaction of nonthermal dark radiation (DR) $A'$ with ordinary photons $A$ via the interaction $eF'_{\mu \nu}F_{\mu \nu}$  \cite{H86}. The DR quanta are much softer, but more numerous than CMB photons,

\begin{equation}
\omega_{DR} \ll \omega_{CMB}, \hspace{.17in} n_{DR}>n_{RJ},  \hspace{.17in} \omega_{DR} n_{DR} \ll \rho_{tot},
\end{equation}
where $\rho_{tot}$ is the total energy density of radiation and DM, $n_{DR}$ is the number density of DR quanta, and $n_{RJ}$ represents the low-energy RJ tail of the standard CMB. Recent papers \cite{CFS15,CPP18} examine interacting DR in the regime where the individual quanta are fewer in number but harder in energy than typical CMB photons, $n_{DR}\ll n_{CMB}$ and $\omega_{DR} \gg \omega_{CMB}$. The interest in an enhancement of the CMB is based on recent tentative observation of a stronger-than-expected absorption signal of 21 cm photons \cite{BRMMM18}, which can be explained by resonant $A'\rightarrow A$ oscillations of dark photons into regular photons in the interval of redshifts $20<z<1700$ \cite{PPRU18}.

This paper is arranged follows. In Sec. II, we review the integral representation for nonthermal or nonequilibrium radiation, where the low-frequency photons are in thermal equilibrium with temperature $T_{RJ}$ owing to bremsstrahlung and the spectrum is bounded from below \cite{MA70,AG80}. In Sec. III, we indicate the upper bound determined by the RJ law with temperature $T_{RJ}$ and obtain the constraints on the photon number per unit volume and the internal energy per unit volume for the CMB that follow from the lower bound of Sec. II. Sec. IV gives a simple illustrative example that shows that the ``heating" of the CMB in the  RJ tail, viz., $T_{RJ}>T_{CMB}$, does actually lead to ``cooling" of the CMB. Finally, Sec. V summarizes our results.

\section{Nonthermal radiation}

The distortions of the CMB has been studied with the aid of the integral representation for the photon number density
\begin{equation}
n_{CMB}(\omega,t) =\frac{1}{\pi^2}\frac{\omega^2}{c^3}\int_{0}^{\infty} \textup{d}T \frac{\sigma(T,t)}{e^{\hbar\omega/k T}-1},
\end{equation}
where the spectral function $\sigma(T,t)$ is positive definite and, in general, is a function of time or the cosmological redshift $z$, where $z\geq 0$ and $z=0$ corresponds to our present time \cite{MA70,AG80}. Such integral representations were obtained  by unifying the notions of the ``approach-to-equilibrium" in quantum statistical mechanics and that of the asymptotic condition in axiomatic quantum field theory in order to describe nonthermal radiation whereby the equilibrium, thermal states form a basis for nonthermal or nonequilibrium states \cite{MA68}.

It is interesting that the distribution (2) appears in an integral equation of Laplace \cite{PW34} by replacing the Maxwell distribution by that of Planck. Paley and Wiener \cite{PW34} describe radiation from a source in approximate local equilibrium by (2), where $\sigma(T,t)$ denotes the ``amount" of radiation coming from black bodies at temperature $T$; consequently, $\sigma(T,t)$ must be positive definite and $\sigma(T,t) \textup{d}T$ gives the amount of radiation coming from the temperature range $T$ to $T+\textup{d}T$.

The case of the integral equation of Laplace \cite{PW34}, viz. Laplace transforms, is the application of the completeness of the equilibrium, thermal states in the description of nonthermal or nonequilibrium states in classical statistical mechanics. The classical integral transform was applied successfully in the study of the approach to equilibrium of Maxwell molecules \cite{MA79}, an exact (similarity) solution of the nonlinear Boltzmann equation \cite{B76,KW76}.  The application of the classical transform, and consequently that for nonthermal radiation was characterized as the temperature integral transform \cite{E81}.

The low-frequency photons comprising the RJ tail of the nonthermal radiation (2) are in thermal equilibrium with temperature
\begin{equation}
T_{RJ}(t)= \int_{0}^{\infty} \textup{d}T \hspace{0.05in} T  \hspace{0.05in} \sigma(T,t),
\end{equation}
which arises from bremsstrahlung processes that are always present.

Spectrum (2) is bounded from below \cite{AG80} by
\begin{equation}
n_{CMB}(\omega,t)\geq \frac{1}{\pi^2}\frac{\omega^2}{c^3} \frac{C(t)}{\exp{[\hbar\omega C(t)/k T_{RJ}(t)]}-1},
\end{equation}
where
\begin{equation}
C(t)=\int_{0}^{\infty} \textup{d}T \hspace{.02in} \sigma(T,t).
\end{equation}
and $C(t)$ and $T_{RJ}(t)$ are, in general, functions of time, viz., redshift  \cite{AG80}.

\section{CMB spectrum constraints}

One obtains an upper bound to  $n_{CMB}(\omega,t)$ with the aid of the inequality $e^x-1 -x \geq 0$ and so (2) implies
\begin{equation}
\frac{1}{\pi^2} \frac{k\omega}{\hbar c^3} T_{RJ}(t) \geq n_{CMB}(\omega,t),
\end{equation}
which bounds the CMB spectrum by the Rayleigh-Jeans law with temperature $T_{RJ}(t)$. This places a constraint on the value of $T_{RJ}(t)$ from obtained data of the cosmological excess at 21 cm emission or absorption signal \cite{BRMMM18}.

One obtains the following inequalities for the number of photons per unit volume $n_{CMB}(t)$ and the internal energy per unit volume $u_{CMB}(t)$ with the aid of inequality (4)
\begin{equation}
n_{CMB}(t) = \int_{0}^{\infty} \textup{d}\omega   \hspace{0.03in}  n_{CMB}(\omega,t) \geq \frac{2 \zeta(3)}{\pi^2(c\hbar)^3}\frac{(kT_{RJ}(t))^3}{(C(t))^2},
\end{equation}
where $\zeta(3)$ is the Riemann's zeta function and
\begin{equation}
u_{CMB}(t) = \int_{0}^{\infty} \textup{d}\omega \hspace{0.02in} \hbar \omega  \hspace{0.02in}  n_{CMB}(\omega,t) \geq \frac{\pi^2}{15(c\hbar)^3} \frac{(kT_{RJ}(t))^4}{(C(t))^3},
\end{equation}
respectively.

The ratio of the internal energy per unit volume $u_{CMB}(t)$ to the internal energy per unit volume $u_{P}(t)$ of a Planckian spectrum with temperature $T_{CMB}(t)$ is
\begin{equation}
\frac{u_{CMB}(t)}{u_{P}(t)} \geq \Big{[}\frac{1}{C(t)}\Big{]}^3 \Big{ [ }\frac{T_{RJ}(t)}{T_{CMB}(t)}\Big{]}^4,
\end{equation}
where
\begin{equation}
n_{P}(\omega,t)=\frac{1}{\pi^2}\frac{\omega^2}{c^3} \frac{1}{e^{\hbar\omega/k T_{CMB}(t)}-1}
\end{equation}
and
\begin{equation}
u_{P}(t)=\int_{0}^{\infty}\textup{d}\omega\hspace{0.03in} \hbar \omega\hspace{0.03in} n_{P}(\omega,t).
\end{equation}
Note, in particular, that if $C(t) =1$ and $T_{RJ} (t)> T_{CMB}(t)$, then the present CMB is ``hotter" than a black-body radiation with temperature $T_{CMB}=2.725 K$, viz., $u_{CMB}(t)> u_{P}(t)$. Therefore, if $T_{RJ}(t) > T_{CMB}(t)$, then one must have $C(t)>1$ to allow the present CMB to be ``cooler" than a black-body radiation with temperature $T_{CMB}$. 

Actually, from (4) one has
\begin{equation}
n_{CMB}(\omega,t)> \frac{1}{\pi^2}\frac{\omega^2}{c^3} \frac{1}{\exp{[\hbar\omega /k T_{CMB}(t)]}-1},
\end{equation}
if $C(t)=1$ and $T_{RJ}(t) > T_{CMB}(t)$ and so there would be a photon number enhancement for all values of $\omega$, which contradicts present CMB data.

\section{Illustrative example}

Recent works suggest a CMB that is very close to a Planckian spectrum but with a significant increase of photon counts in the RJ tail \cite{PPRU18}. We present a simple example of the distortions of the CMB, from that of a pure Planckian spectrum with temperature $T_{CMB}$, that follows from the integral representation (2).
Consider
\begin{equation}
u_{CMB}(\omega)= \frac{\omega^3}{\pi^2}\Big{[}\frac{C_{1}}{e^{\omega/T_{1}}-1}  +\frac{C_{2}}{e^{\omega/T_{2}}-1}\Big{]},
\end{equation}
where  we have chosen $\hbar =c=k=1$. One has from (3) and (5) that
\begin{equation}
C= C_{1}+ C_{2}\hspace{0.15in} \textup{and} \hspace{0.15in}  T_{RJ}= C_{1}T_{1}+ C_{2} T_{2}.
\end{equation}
Let $C_{1}=C_{2}=1$ and $T_{1}=0.999 \hspace{0.03in} T_{CMB}$ and $T_{RJ}=1.060 \hspace{0.03in} T_{CMB}$, which implies that $T_{2}=0.061 T_{CMB}$.

Fig. 1 shows the plots for the upper bound given in (6) (green), the lower bound in (4) (red), our example in (13) (blue), and the Planckian black body spectrum in (12) (black). The present temperature of the CMB is $T_{CMB}=2.725 K$. Note that the blue plot is between the red and the blue, which will always be the case. The dot in the green plot at $\omega/T_{CMB} =0.0251$ represents the upper bound (6) for the 21 cm photons for $T_{RJ}=1.060  T_{CMB}$. Note that the observation of the absorption signal of 21 cm photons will determine the maximum possible value of $T_{RJ}$.

\begin{figure}
\begin{center}
   \includegraphics[scale=0.3]{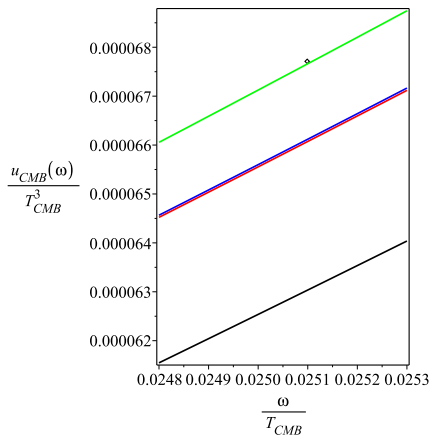}
\end{center}
\label{fig:theFig}
  \caption{Plots of the upper bound given in (6) (green), the lower bound in (4) (red), our example in (13) (blue) and the Planckian black body spectrum in (12) (black) at the temperature of the CMB as observed in the present day, viz., $T_{CMB}=2.725 K$. The dot in the green plot at $\omega/T_{CMB} =0.0251$ represents the upper bound (6) for the 21 cm photons for $T_{RJ}=1.060  T_{CMB}$.}
\end{figure}

Fig. 2 shows plots for our example in (13) and the Planckian black body spectrum in (12) (black). The plots cross at $\omega/T_{CMB} = 0.360$ (not shown) with the higher values of the blue plot over the black plot representing the enhancement of photons over that of the black body radiation. The enhancement of photons in the RJ tail does not give rise to an increase in the internal energy of our example (13) over that of the black body radiation given in (12). In fact, the internal energy per unit volume of both spectra are about the same since the contribution of the spectrum with temperature $T_{2}$ in (13) is negligible in comparison to the overall internal energy per unit volume.

\begin{figure}
\begin{center}
   \includegraphics[scale=0.3]{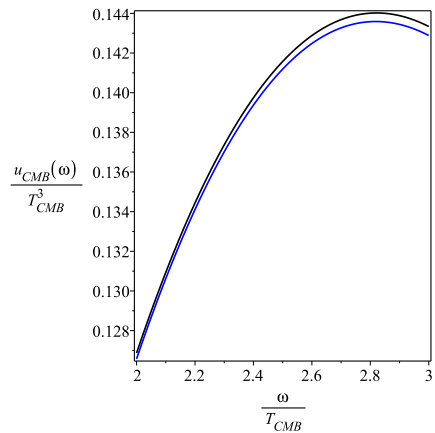}
\end{center}
\label{fig:theFig}
  \caption{The blue plot represents our example (13) and the black that of the black body radiation in (12). The plots cross at $\omega/T_{CMB} = 0.360$ (not shown) with the higher values of the blue plot over the black plot representing the enhancement of photons over that of the black body radiation with temperature $T_{CMB}$.}
\end{figure}

Fig. 3 shows the region of maximum difference between our example (13) and the black body radiation in inequality (12). The difference between the two plots for $T_{RJ}= 1.060 \hspace{0.03in} T_{CMB}$ is $0.03\%$ at $x=2.82$, where the black body radiation attains its maximum value.
\begin{figure}
\begin{center}
   \includegraphics[scale=0.3]{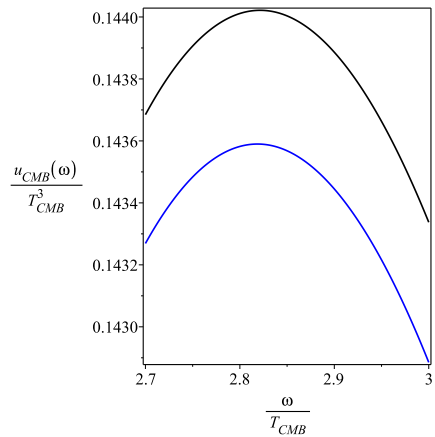}
\end{center}
\label{fig:theFig}
  \caption{Region of maximum difference between our example (13) and the black body radiation in inequality (12). }
\end{figure}
Therefore, the model of interacting DR quanta much softer, but more numerous than CMB photons \cite{PPRU18}, may result in actually ``cooling" the CMB radiation even though $T_{RJ}>T_{CMB}$ implies the enhancement of photons in the RJ tail, viz., the ``heating-up" of low frequency photons.

It is important to remark that $C(t)=1$ is in total disagreement with the data for the CMB. The lower bound (4) implies for $T_{RJ}= 1.060 \hspace{0.03in} T_{CMB}$, that is, a $6\%$ increase of the temperature $T_{RJ}$ over $T_{CMB}$, that at x=2.82, which is near the peak of the black body radiation, the lower bound (4) with $C(T)=1$ is $19\%$ higher than the black body radiation and so in total disagreement with the data. This, of course, implies that $C(t)>1$, which is the case for our example (13) of the two-temperature approximation to the integral  (2), and indicates an additional cosmological contribution to the present CMB besides the contribution of the original remnant of the Big Bang.

\section{Summary and conclusion}

We have established several constraints on the spectrum of the CMB that follow from the integral representation (2). One obtains an upper bound (6) to the spectrum given by the Rayleigh-Jeans law with temperature $T_{RJ}$ and a lower bound given by (4) \cite{MA70,AG80}. Observations of the absorption signal of the 21 cm photons set an upper bound to the value of $T_{RJ}$. Also, if $T_{RJ} > T_{CMB}$, then we must have that $C(t)>1$ in (5) which requires additional cosmological sources which when added to the remnant radiation from the Big Bang gives us the present CMB. We believe that the integral representation (2) for nonthermal radiation may be considered as an additional constraint to other cosmological or astrophysical constraints in the study of the distortions of the CMB brought about by new physics particles or fields.

\bibliography{basename of .bib file}

\end{document}